\documentclass[12pt,a4paper]{article}
\usepackage{amsmath}
\usepackage[dvips]{graphicx}
\input epsf
\usepackage{times}
\begin{document}
\begin{titlepage}
\title{Transient state of matter in hadron and nucleus collisions}
\author{S.M. Troshin,
 N.E. Tyurin\\[1ex]
\small  \it Institute for High Energy Physics,\\
\small  \it Protvino, Moscow Region, 142281, Russia}
\normalsize
\date{}
\maketitle

\begin{abstract}
We discuss properties of the specific strongly interacting
transient collective state of matter in hadron and nuclei
reactions and emphasize  similarity in their dynamics. We consider
elliptic flow introduced for description of nucleus collisions
and discuss its possible behavior  in hadronic reactions due to rotation
of the transient matter.
 \\[2ex]
 PACS: 12.38.Mh, 21.60.Ev\\
 Keywords: transient state,
hadron interactions,  collective effects of rotation,
elliptic flow

\end{abstract}
\end{titlepage}
\setcounter{page}{2}

\section*{Introduction}
Multiparticle production in hadron and nucleus collisions and
corresponding observables  provide  a clue to the mechanisms of
confinement and hadronization. Discovery of the deconfined state
of matter has been announced  by four major experiments at RHIC
\cite{rhic}. Despite the highest values of energy and density have
been reached, a genuine quark-gluon plasma QGP (gas of the free
current quarks and gluons) was not found. The deconfined state
reveals the properties of the perfect liquid, being strongly
interacting collective state and therefore it was labelled as sQGP
\cite{denteria}. These results immediately have became a subject
for an  active theoretical studies. The nature of this new form of
matter is not known and the variety of models has been proposed to
treat its properties \cite{models}. The importance of this result
is that the matter is still strongly correlated
and reveals high degree of the coherence when it is
 well beyond the critical values of density and
temperature. The elliptic flow and constituent quark scaling of the
observable $v_2$ demonstrated  an importance of the constituent quarks
 \cite{volosh} and their role as
effective degrees of freedom of the newly discovered form of matter.
Generally speaking this result has shown  an importance of the nonperturbative
effects in the region where such effects were  not expected.
Review paper  which provides
an emphasis on the historical aspects of the QGP searches was published in \cite{weiner}.
The important conclusion made in this paper is that the  deconfined state
of matter has already been observed in hadronic reactions and it would be interesting
to study collective properties of transient state in reactions with hadrons and nuclei
simultaneously.

In this paper we also note that the behavior of collective observables
 in hadronic and nuclear
reactions could have similarities. We discuss the role of the coherent rotation
of the transient matter in hadron and nuclei reactions and dependence of the anisotropic
flows.

\section{Experimental probes of collective dynamics. Constituent quark scaling.}

There are several experimental probes  of collective dynamics in
$AA$ interactions \cite{voloshin,molnar}. A  most widely discussed one
is the elliptic flow
\begin{equation}\label{v2}
v_2(p_\perp)\equiv \langle \cos(2\phi)\rangle_{p_\perp}=\langle \frac{p_x^2-p_y^2}{p_\perp^2}\rangle,
\end{equation}
which is the second Fourier moment of the azimuthal momentum distribution of particles
 at fixed value of $p_\perp$.
The common origin of the elliptic flow is considered to be an  almond shape of
the overlap region of the two spherically symmetrical colliding
nuclei and strong interaction in this region. The azimuthal angle $\phi$ is
the angle of the detected particle with respect to the reaction
plane, which is spanned by the collision axis $z$ and the impact parameter vector $\mathbf b$. The impact
parameter vector $\mathbf b$ is directed along the $x$ axis. Averaging is taken over large number
 of the events. Elliptic flow can be expressed  in covariant form in
 terms of the impact parameter and transverse momentum
 correlations as follows
 \begin{equation}\label{v2a}
v_2(p_\perp)=\langle \frac{(\hat{\mathbf  b}\cdot {\mathbf  p}_\perp)^2}{p_\perp^2}\rangle-
\langle\frac{(\hat{\mathbf  b}\times {\mathbf  p}_\perp)^2}{p_\perp^2}\rangle ,
\end{equation}
where $\hat{\mathbf  b}\equiv \mathbf  b /b$. In more general
terms, the momentum anisotropy $v_n$ can be characterized
according
 to the Fourier expansion of the freeze-out source distribution \cite{molnar}:
 \begin{equation}\label{sor}
S(x,\mathbf{p}_\perp,y)\equiv dN/d^4xd^2p_\perp dy
\end{equation}
in terms of the momentum azimuthal angle.

The observed elliptic flow $v_2$ is the weighted
average of $v_2(x,p_\perp,y)$ defined in the infinitesimal spacetime volume $d^4x$.
Common explanation of the dynamical origin of elliptic flow is the strong scattering during the early
stage of interaction in the overlap region.

There is an extensive set of the experimental data for the elliptic flow $v_2$
in nucleus-nucleus collisions (see for the recent review, e.g. \cite{lacey}).
Integrated elliptic flow $v_2$ has a nontrivial  dependence on $\sqrt{s_{NN}}$:
at low energies it demonstrates sign-changing behavior, while at high energies
$v_2$ is positive and increases with $\sqrt{s_{NN}}$ linearly.

The differential elliptic flow $v_2(p_\perp)$ increases with $p_\perp$
at small values of transverse
momenta, then it becomes flatten in the region of the intermediate transverse
momenta and decreases at large $p_\perp$, but to a non-zero value. The magnitude
of $v_2$ in the region of intermediate $p_\perp$ is rather high at RHIC
and has a value about $0.2$ close to hydrodynamical limit \cite{lacey} indicating
presence of order and pair correlations relevant for the liquid phase. The increase
of elliptic flow at small transverse momenta is in a good agreement with hydrodinamical
model while the experimental data deviate from this model at higher values of transverse
momenta \cite{hydro}.

An interesting property
of the differential elliptic flow $v_2(p_\perp)$ in $AA$-collisions --- the constituent
 quark scaling \cite{volosh}. We  discuss
 it in a some  detail now. The scaling occurs if hadronization mechanism
goes via coalescence of the constituent quarks and it is expressed as an approximate relation
 $v_2(p_\perp)\simeq n_Vv_2(p_\perp/n_V)$, where $n_V$ is the number of the valence constituent quarks
 in the hadron. This scaling takes place in the region of the intermediate transverse
  momenta and reveals important role of constituent quarks in the deconfined phase
reached in nucleus collisions \cite{eremin}. The quantity $v_2/n_V$ can be interpreted
as an elliptic flow of a constituent quark $v^{Q}_{2}$. It increases with transverse
momentum in the region $0 \leq p^{Q}_{\perp} \leq 1$ $GeV/c$ and with a rather
good accuracy does not depend on $p^{Q}_{\perp}$
at $p^{Q}_{\perp}\geq 1$ $GeV/c$.

In the following section we will discuss energy and transverse momentum dependencies of $v_2$
  in hadron collisions at
 fixed impact parameters and extend this consideration for nucleus
collisions with emphasis on the similarity  of the transient states in hadron
and nucleus collisions. We  consider non-central hadron collisions
and apply notions acquired from heavy-ion studies. It is reasonable to do so in the framework
of the constituent quark model picture for hadron structure where hadrons look similar to the
light nuclei. In particular, we  amend the model \cite{csn} for hadron interactions based on the chiral quark
model ideas  and consider the effect of collective rotation of the quark matter in the overlap region.
All that was said above might have a particular interest under studies of hadron collisions in the new
few TeV energy region where number of  secondary particles will increase
 significantly indicating importance of collective effects.

\section{Transient state of matter in hadron collisions}
In principle, the geometrical picture of hadron collision is in complete analogy
with nucleus collisions and we believe that the assumption \cite{polvol} on the possibility to
determine reaction plane in the non-central hadronic collisions can be
 justified experimentally and the standard procedure\cite{poskanz} can be used.
It would be useful to perform the measurements of the characteristics of
 multiparticle production processes in hadronic collisions
at fixed impact parameter  by  selecting
specific events sensitive to its value and direction.
The relationship of the impact parameter with the
 final state multiplicity is a useful tool in these studies similar to the studies
 of the nuclei interactions, e.g in the Chou-Yang approach  \cite{chyn} one can restore the
 values of impact parameter from the charged particle multiplicity
\cite{chmult}.
Thus, the impact parameter can be  determined through
 the centrality \cite{bron} and then, e.g. elliptic flow, can be analyzed selecting events
 in a specific centrality ranges. Indeed, in the work \cite{bron} the following relation
 \begin{equation}\label{cent}
 c(N)\simeq \frac{\pi b^2(N)}{\sigma_{inel}},
\end{equation}
for the values of the impact parameter $b< \bar{R}$ can be
 extended straightforwardly to the case of hadron scattering.
 Then we should consider $\bar{R}$ as a sum of the two radii of colliding hadrons
 and $\sigma_{inel}$ as the total inelastic hadron-hadron cross--section. The centrality
 $c(N)$ is the centrality of the events with the multiplicity larger than $N$ and $b(N)$ is
 the impact parameter where the mean multiplicity $\bar n (b)$ is equal to $N$.
 The centrality can be determined by the fraction of the events with the largest number of
 produced particles which are registered by detectors \cite{bron,antinori}.

Of course, the standard inclusive cross-section for unpolarized
particles being integrated over impact parameter $\mathbf b $,
cannot depend on the azimuthal angle of the detected particle
transverse momentum. We need to be a more specific at this
point and consider for discussion of the azimuthal angle dependence
 some particular form for the inclusive
 cross-section.
 For example, with account for
$s$--channel unitarity  inclusive cross-section  can be written in the following form
\begin{equation}
\frac{d\sigma}{d\xi}= 8\pi\int_0^\infty
bdb\frac{I(s,b,\xi)}{|1-iU(s,b)|^2}\label{unp}.
\end{equation}
Here the function $U(s,b)$ is similar to an input Born amplitude  and
related to the elastic scattering scattering amplitude through an algebraic
equation which enables one to restore unitarity \cite{umat}. The set of kinematic variables denoted
by $\xi$ describes the state of the  detected particle.
This function is constructed from the multiparticle analogs $U_n$ of the function $U$
and is in fact an
inclusive cross-section in the impact parameter space without account for the unitarity corrections, which
are given by the factor
\[
w(s,b)\equiv |1-iU(s,b)|^{-2}
\]
 in Eq. (\ref{unp}).
Unitarity, as it will be evident from the following, modifies anisotropic flow.
When the impact parameter vector $ \mathbf {b}$ and transverse momentum ${\mathbf  p}_\perp $
of the detected particle are fixed,
the function $I=\sum_{n \geq 3} I_n$, where $n$ denotes  a number of particles in the final state,
  depends on the azimuthal angle $\phi$ between
 vectors $ \mathbf b$ and ${\mathbf  p}_\perp $.
It should be noted that the impact parameter
$ \mathbf {b}$ is the  variable conjugated to the transferred momentum
$ \mathbf {q}\equiv \mathbf {p}'_a-\mathbf {p}_a$ between two incident channels
 which describe production processes
of the same final multiparticle state.
The dependence on the azimuthal angle $\phi$ can be written in explicit form through the Fourier
series expansion
\begin{equation}\label{fr}
I(s,\mathbf b, y, {\mathbf  p}_\perp)=\frac{1}{2\pi}I_0(s,b,y,p_\perp)[1+
\sum_{n=1}^\infty 2\bar v_n(s,b,y,p_\perp)\cos n\phi].
\end{equation}
The function $I_0(s,b,\xi)$ satisfies  to the
following sum rule
\begin{equation}\label{sumrule}
\int I_0(s,b,y,p_\perp) p_\perp d p_\perp dy=\bar n(s,b)\mbox{Im} U(s,b),
\end{equation}
where $\bar n(s,b)$ is the mean multiplicity depending on impact parameter.
Thus, the bare  flow $\bar v_n(s,b,y,p_\perp)$ is related to the
measured  flow $v_n$  as follows
\[
v_n(s,b,y,p_\perp)=w(s,b)\bar v_n(s,b,y,p_\perp).
\]
In the above formulas the variable $y$ denotes rapidity, i.e. $y=\sinh^{-1}(p/m)$,
where $p$ is a longitudinal momentum.
Thus, we can see that unitarity corrections are mostly important
at small impact parameters, i.e. they modify flows at small centralities,
while peripheral collisions are almost not affected by unitarity.
The following limiting behavior of $v_n$ at $b=0$ can be easily obtained:
\[
v_n(s,b=0, y,p_\perp)\to 0
\]
at $s\to\infty$ since $U(s,b=0)\to\infty$ in this limit.

General considerations demonstrate that we could expect significant
values of directed $v_1$ and elliptic  $v_2$ flows in hadronic interactions.
 For example,
according to the uncertainty principle we can estimate the value
of $p_x$ as $1/\Delta x$ and correspondingly $p_y\sim 1/\Delta y $
where $\Delta x$ and $\Delta_y$ characterize the size of the
region where the particle originate from. Taking $\Delta x \sim
R_x$ and $\Delta_y \sim R_y$, where $R_x$ and $R_y$ characterize
the sizes of the almond-like overlap region in transverse plane,
 we can easily
obtain proportionality of $v_2$ to the eccentricity of the overlap
region, i.e.
\begin{equation}\label{exc}
v_2(p_\perp)\sim \frac{R_y^2-R_x^2}{R_x^2+R_y^2}.
\end{equation}
The presence of correlations of impact parameter vector $\mathbf
b$ and $\mathbf p_\perp$ in hadron interactions follows also from the relation between
impact parameters in the multiparticle production\cite{webb}:
\begin{equation}\label{bi}
{\mathbf b}=\sum_i x_i{ \tilde{\mathbf  b}_i}.
\end{equation}
Here  $x_i$ stand for Feynman $x_F$ of $i$-th particle, the impact
parameters $\tilde {\mathbf b}_i$ are conjugated to the transverse
momenta $\tilde {\mathbf p}_{i,\perp}$. Such correlation should be
more prominent in the large-$x_F$ (fragmentation) region\footnote{It
should be noted that the directed flow
$v_1(p_\perp)\equiv \langle \cos \phi \rangle_{p_\perp}=\langle
{\hat{\mathbf  b}\cdot {\mathbf  p}_\perp} /{p_\perp}\rangle
$
 the measurements at RHIC \cite{dir} are in agreement with the above conclusion.}.

The above  considerations are  based on the uncertainty principle and angular momentum
conservation, but they do not preclude an existence of the dynamical description in the
terms similar to the ones used in heavy-ion collisions, i.e.
the underlying  dynamics  could be  the same as the dynamics of the elliptic flow in
nuclei collisions and transient state can originate from the
nonperturbative sector of QCD.

We would like to point out to the possibility that
the transient state in both cases  can be related to
the mechanism of spontaneous chiral symmetry breaking ($\chi$SB) in QCD \cite{bjorken},
 which  leads
to the generation of quark masses and appearance of quark condensates. This mechanism describes
transition of current into  constituent quarks, which are
   the quasiparticles with masses
 comparable to  a hadron mass scale.
The  gluon field is responsible for providing quarks
  masses and internal structure through the instanton
  mechanism of the spontaneous chiral symmetry
 breaking \cite{inst}.

  Collective excitations of the condensate are the Goldstone bosons
and the constituent quarks interact via exchange
of the Goldstone bosons; this interaction is mainly due to a pion field\cite{diak}.
The  general form of the effective Lagrangian (${\cal{L}}_{QCD}\rightarrow {\cal{L}}_{eff}$)
 relevant for
description of the non--perturbative phase of QCD proposed in \cite{gold}
 and includes the three terms \[
{\cal{L}}_{eff}={\cal{L}}_\chi +{\cal{L}}_I+{\cal{L}}_C.\label{ef} \]
Here ${\cal{L}}_\chi $ is  responsible for the spontaneous
chiral symmetry breaking and turns on first.  To account for the
constituent quark interaction and confinement the terms ${\cal{L}}_I$
and ${\cal{L}}_C$ are introduced.  The  ${\cal{L}}_I$ and
${\cal{L}}_C$ do not affect the internal structure of the constituent
quarks.

The picture of a hadron consisting of constituent quarks embedded
 into quark condensate implies that overlapping and interaction of
peripheral clouds   occur at the first stage of hadron interaction.
At this stage the part of the effective lagrangian ${\cal{L}}_C$ is turned off
(it is turned on again in the final stage of the reaction).
Nonlinear field couplings   transform then the kinetic energy to
internal energy and mechanism of such transformations was discussed
 by Heisenberg \cite{heis} and  Carruthers \cite{carr}.
As a result the massive
virtual quarks appear in the overlapping region and  some effective
field is generated. This field is generated by $\bar{Q}Q$ pairs and
pions strongly interacting with quarks. Pions themselves are the bound states of massive
quarks. This part of interaction is described by ${\cal{L}}_I$ and
a possible form of ${\cal{L}}_I$ was discussed in \cite{diakp}.

The generation time of the effective field (transient phase) $\Delta t_{eff}$
\[
\Delta t_{eff}\ll \Delta t_{int},
\]
where $\Delta t_{int}$ is the total interaction time. This assumption on the almost instantaneous
generation of the effective field has obtained support in the very short thermalization time revealed
in heavy-ion collisions at RHIC \cite{therm}.

Under construction of particular model \cite{csn} for the function
$U(s,b)$ it was supposed that the valence quarks located in the
central part of a hadron were scattered in a
quasi-independent way by the effective field.
In accordance with the
quasi-independence of valence quarks  the basic
dynamical  quantity is represented in  the form  of  the product \cite{csn} of factors
$\langle f_{Q}(s,b)\rangle$ which correspond to the individual quark
scattering
 amplitudes which are integrated  over transverse position distribution of $Q$
 inside its parent hadron
  and  the longitudinal momentum distribution  carried by quark $Q$.
The integrated amplitude $\langle f_Q(s,b)\rangle $ describes averaged elastic
   scattering  of a single
valence   quark $Q$ in the effective field, its  interaction radius is determined
by the quark mass:
\begin{equation}\label{rq}
R_Q=\xi/m_Q.
\end{equation}
Factorization in the impact parameter representation reflects the coherence in the valence
quark scattering,  it corresponds to the simultaneous
scattering of valence quarks by the effective field. This mechanism
resembles  Landshoff mechanism of the simultaneous quark--quark independent
  scattering \cite{landso}.  However, in our case we suppose validity of
the  Hartree--Fock approximation for the
constituent quark scattering in the mean field. Thus, $U$-matrix is a product
of the averaged single quark scattering amplitudes, but the resulting $S$-matrix
cannot be factorized and therefore the term quasi-independence is relevant.
The above picture assumes  deconfinement at the initial stage of
 the hadron collisions and  generation of common for both hadrons mean field during the first stage.
Those notions were  used in the model \cite{csn} which has
been applied to description of elastic scattering. Here we will extend them to particle
production with account of the geometry of the region where the effective field (quarks interacting
by pion exchange) is located and conservation of angular momentum.

To estimate the number
of scatterers in the effective field one could assume that  part of hadron energy carried by
the outer condensate clouds is being released in the overlap region
 to generate massive quarks. Then this number can be estimated  by:
 \begin{equation} \tilde{N}(s,b)\,\propto
\,\frac{(1-\langle k_Q\rangle)\sqrt{s}}{m_Q}\;D^{h_1}_c\otimes D^{h_2}_c
\equiv N_0(s)D_C(b),
\label{Nsbt}
\end{equation} where $m_Q$ -- constituent quark mass, $\langle k_Q\rangle $ --
average fraction of
hadron  energy carried  by  the valence constituent  quarks. Function $D^h_c$
describes condensate distribution inside the hadron $h$, and $b$ is
an impact parameter of the colliding hadrons.
 In elastic scattering the massive
virtual  quarks are transient
ones: they are transformed back into the condensates of the final
hadrons.
The overlap region, which described by the function $D_C(b)$,
has an ellipsoidal form similar to the overlap region in the nucleus collisions (Fig. 1).
\begin{figure}[hbt]
\begin{center}
\epsfxsize=  70 mm  \epsfbox{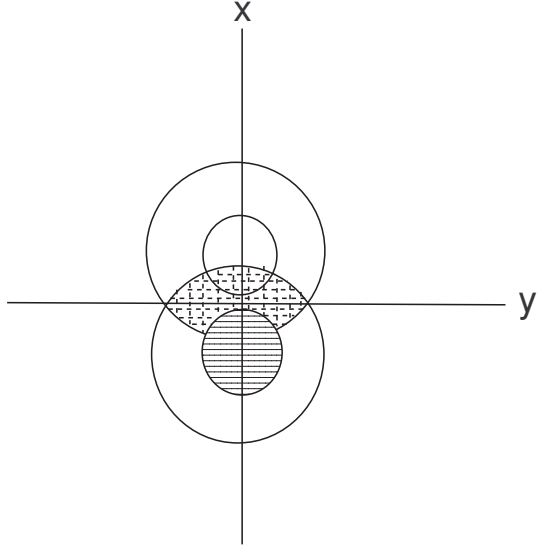}
\end{center}
\caption{Schematic view in frontal plane of the hadron collision as extended objects.
Collision occurs along the z-axis.}
\end{figure}

Valence constituent quarks would excite a part of the cloud of the virtual massive
quarks and those
quarks will subsequently hadronize  and form the multiparticle
final state.  Existence of the massive quark-antiquark matter in the stage
preceding
hadronization seems to be
supported  by the experimental data obtained
at CERN SPS and RHIC (see \cite{biro} and references therein)

The  geometrical picture of hadron collision discussed above
implies that the generated massive
virtual  quarks in overlap region carries large orbital angular momentum
at high energies and non-zero impact parameters. The total orbital angular
momentum  can be estimated
as follows
\begin{equation}\label{l}
 L \simeq \alpha b \frac{\sqrt{s}}{2}D_C(b),
\end{equation}
where parameter $\alpha$ is related to the fraction of the initial energy carried by the condensate
clouds which goes to rotation of the quark system.
Due to strong interaction
between quarks this orbital angular momentum  leads to the coherent rotation
of the quark system located in the overlap region as a whole  in the
$xz$-plane (Fig. 2). This rotation is similar to the liquid rotation
where strong correlations between particles momenta exist.
This point is different from the parton picture used in \cite{wang},
where collective rotation of a parton system as a whole
was not anticipated.
\begin{figure}[hbt]
\begin{center}
\epsfxsize=  70 mm  \epsfbox{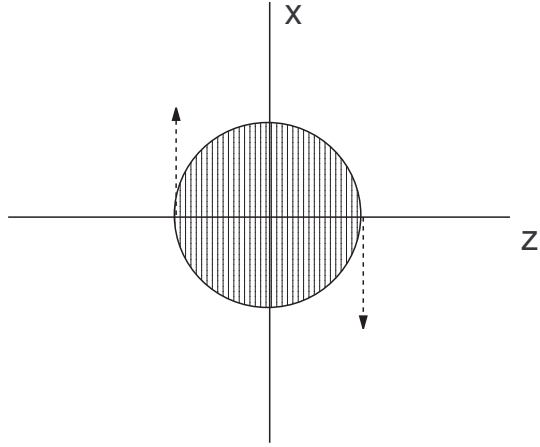}
\end{center}
\caption{Collective rotation of the overlap region, view in the $xz$-plane.}
\end{figure}
This is a main point of the proposed mechanism of the elliptic flow in hadronic collisions
 --- collective rotation
of the strongly interacting system of massive
virtual quarks. Number of the quarks in this system
is proportional to $N_0(s)$ and it is natural to expect therefore that the integrated elliptic flow
$v_2\propto \sqrt{s}$. Such dependence of $v_2$ is in a good agreement with experimental
data for nucleus collisions and this implies already mentioned similarity between hadron and nucleus
reactions. The same origin, i.e. proportionality to the  quark number in
 the transient state, has the preasymptotic increase of the total cross-sections \cite{nadol}.
\begin{equation}
 \sigma_{tot}(s)=a+b\sqrt{s}
 \end{equation}
 in the region up to $\sqrt{s}\sim 0.5$ $TeV$. At higher energies unitarity transforms
 such dependence into $\ln^2 s$.

We  consider  now effects of rotation for the differential elliptic flow $v_2(p_\perp)$.
We  would like to recall  that the assumed particle production mechanism is
the excitation of  a part of the rotating cloud of the virtual massive constituent
quarks by the one of the valence constituent quarks with  subsequent hadronization.

 Different
mechanisms of the hadronization will be discussed later, and now we will concentrate on
the differential elliptic flow $v^Q_2(p_\perp)$ for constituent quarks. It is natural to
suppose that the size of the region where the virtual massive quark $Q$ is knocked out from the cloud
is determined by its transverse momentum, i.e. $\bar R\simeq 1/p_\perp$. However, it is
evident that $\bar R$ cannot be larger than the interaction radius of the valence
 constituent quark $R_Q$ which
 interacts with the massive
virtual quarks  quarks from the cloud.
It is also clear that $\bar R$ cannot be less than the geometrical size of the valence constituent
quark $r_Q$. The magnitude of the quark interaction radius was
 obtained under analysis of elastic scattering \cite{csn}
 and has the following dependence on the valence constituent quark mass in the form (\ref{rq}),
where $\xi \simeq 2$ and therefore $R_Q\simeq 1$ $fm$, while the geometrical radius of  quark $r_Q$
is about $0.2$ $fm$.
The size of the region\footnote{For simplicity we suppose that this region has a spherically
symmetrical form} which is responsible for the small-$p_\perp$ hadron production is large,
valence constituent quark excites rotating cloud of quarks with various values and directions
of their momenta in that case. Effect of rotation will be smeared off in the volume $V_{\bar R}$
 and therefore $\langle \Delta p_x \rangle_{V_{\bar R}} \simeq 0$ (Fig. 3, left panel). Thus,
\begin{equation}
\label{larg}
v^Q_2(p_\perp)\equiv\langle v_2\rangle_{V_{\bar R}}\simeq 0
\end{equation}
at small $p^Q_\perp$.
\begin{figure}[hbt]
\begin{center}
\epsfxsize=  40 mm  \epsfbox{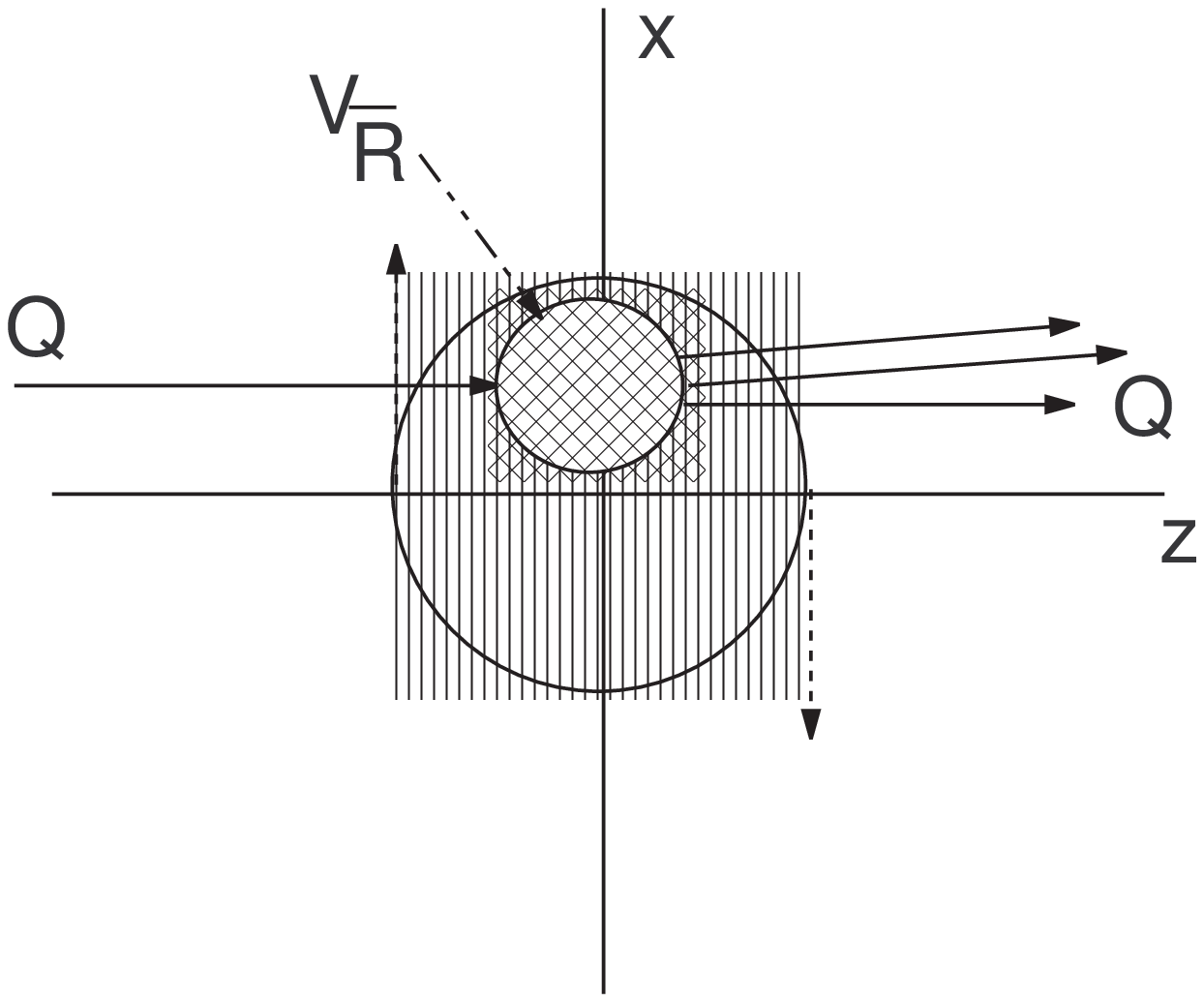}\,\quad\quad\,
\epsfxsize=  40 mm  \epsfbox{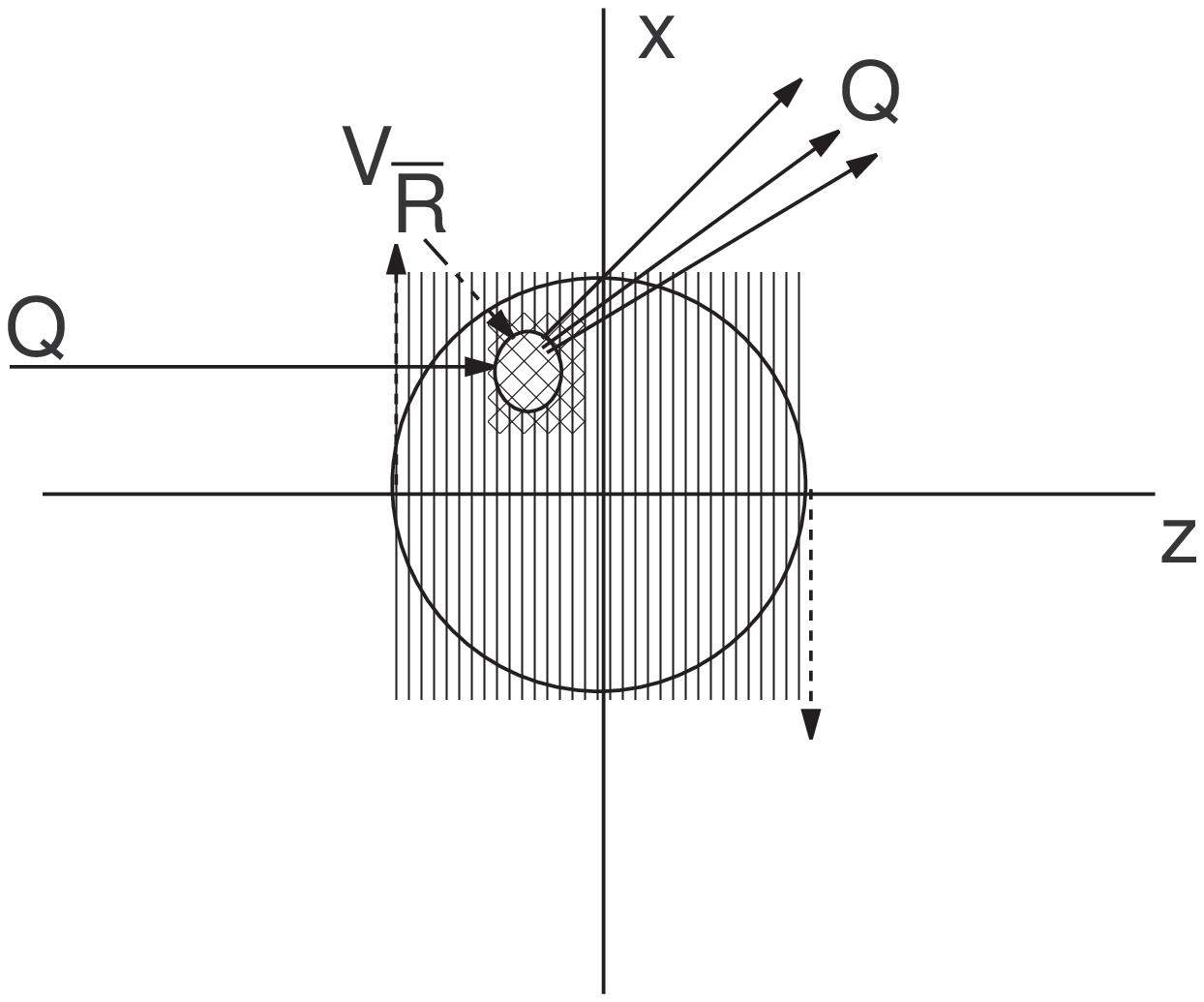}
\end{center}
\caption{Schematic view of the rotation effect in the production of constituent quarks
with small $p^Q_\perp$ (left panel) and with large $p^Q_\perp$ (right panel),
 view in the $xz$-plane.}
\end{figure}
When we proceed to the region of higher values of $p^Q_\perp$, the radius $\bar R$ is decreasing
and the  effect of rotation  becomes more and more prominent, valence quark excites now the region
where most of the quarks move coherently, i.e. in the same direction, with approximately
the same velocity
(Fig. 3, right panel). The mean value $\langle \Delta p_x \rangle_{V_{\bar R}} > 0$ and
\begin{equation}
\label{smal}
v^Q_2(p_\perp)\equiv\langle v^Q_2\rangle_{V_{\bar R}}> 0
\end{equation}
and increase with increasing $p_\perp$.
However, as is was already mentioned $\bar R$ cannot be smaller than the geometrical radius
of constituent quark and therefore the increase of $v^Q_2$ with $p^Q_\perp$ will disappear when
$\bar R =r_Q$, i.e. at $p^Q_\perp \geq 1/r_Q$, and saturation will take place.
The value of transverse
momentum where the saturation starts is about $1$ $GeV/c$ for $r_Q\simeq 0.2$ $fm$.
Thus, the qualitative dependence of $v^Q_2(p_\perp)$\footnote{It is worth to note that the subscript
$Q$ is used for the incoming constituent quark, while the superscript $Q$ being used for
the outgoing constituent quarks} will have a form depicted in Fig. 4.
\begin{figure}[hbt]
\begin{center}
\epsfxsize=  70 mm  \epsfbox{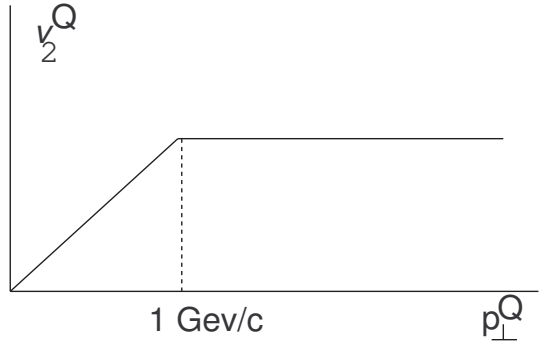}
\end{center}
\caption{Qualitative dependence of the elliptic flow $v^Q_2$ of constituent quarks
on transverse momentum.}
\end{figure}

 Predictions for the elliptic flow for the particular
hadron depends on the supposed mechanism of hadronization. For the
region of the intermediate values of $p_\perp$ the constituent
quark coalescence mechanism \cite{volosh} would be dominating one.
In that case values for hadron elliptic flow can be obtained from
the constituent quark one by the replacement $v_2\to n_Vv^Q_2$ and
$p_\perp\to p^Q_\perp/n_V$.

 However,
the fragmentation mechanism should also be present at small and large
transverse momenta, and it will survive at large $p_\perp$.
 As a possible choice for the fragmentation
mechanism, the   chiral quark models can be used and it implies that
the virtual massive quark $Q$
fluctuates into Goldstone boson and
  another constituent quark $Q'$  \cite{cheng}:
\begin{equation}\label{trans}
Q\to GB+Q',
\end{equation}
where $GB$ denotes Goldstone bosons (Fig. 5).
\begin{figure}[h]
\begin{center}
  \resizebox{4cm}{!}{\includegraphics*{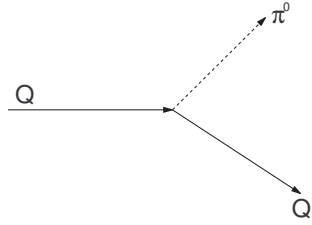}}
\end{center}
\caption{Schematical view of the quark fragmentation into $\pi^0$ in the chiral quark models.
 \label{ts1}}
\end{figure}

Elliptic flow of the  quarks and elliptic flow of the
hadron are approximately equal for the fragmentation process . Thus, in the region of the
intermediate transverse momenta
 elliptic flow of quarks will be enhanced due to quark coalescence and at higher
 transverse momenta the elliptic flow will level off and return to the flat dependence
 of the quark elliptic flow (Fig. 4).

The considered mechanism of particle production has a two-step nature and based on the independent
excitation of the rotating cloud by the valence quarks. It  would lead therefore to the negative
 binomial form of the multiplicity distribution
\begin{equation}\label{md}
P_n=\frac{\Gamma(n+k)}{\Gamma(n+1)\Gamma(k)}\left[\frac{\bar n}{\bar n+k}\right]^n\left[\frac{k}{\bar n+k}\right]^k,
\end{equation}
where parameter $k=\langle N\rangle $, i.e. it should be interpreted
 as the averaged over impact parameter number of the active valence quarks (quarks which excite the cloud)
at the given high energy:
\begin{equation}\label{kp}
\langle N\rangle=\frac{\int_0^\infty bdb N(b)\eta(s,b)}{\int_0^\infty bdb \eta(s,b)},
\end{equation}
where $N(b)$ is the distribution of the quark number over impact parameter and
\[
\eta(s,b)\equiv\frac{1}{4\pi}\frac{d\sigma_{inel}}{db^2}
\]
is the inelastic overlap function. Since we adopted hadron structure with the valence constituent
quarks in the central part, the function $N(b=0)=N$, where $N=n_{h_1}+n_{h_2}$ is the total number
of the valence quarks in the colliding hadrons.  Eq. (\ref{kp}) implies  that the probability of
inelastic interaction of valence quark is proportional to the probability of the hadron inelastic interaction.
 The form (\ref{md}) for the multiplicity distribution is in a good agreement
with experimental data, e.g. at CERN SPS energy, where parameter $k$ varies in the region from 4.6 to 3.2
\cite{chyn}. The parameter $k$ is decreasing with energy.

 It would be useful to asses other effects of the proposed mechanism, where rotation of cloud of the
 virtual massive quarks appears as a main point. Due to this rotation the density of  massive quarks will be
 different in the different parts of the cloud, it will be smaller in the central part and bigger at
 the peripheral part of cloud due to the centrifugal effect. At the same time the quarks
 in the peripheral part have a maximal transverse momenta and therefore we should observe correlation
 of the multiplicity and transverse momentum. Indeed, in the assumed mechanism of particle production
 with large
 transverse momenta ($p_\perp \geq 1$ GeV/c) the interaction region of valence constituent quark
  with the cloud is determined
 by the geometrical radius of the quark $r_Q$ and therefore the mean associated multiplicity at fixed impact
 parameter $\bar n(s,\mathbf b, {\mathbf p}_\perp)$ should increase with $p_\perp$ and depend on the azimuthal
  angle between vectors $\mathbf b$ and ${\mathbf p}_\perp$. It should be noted that
\begin{equation}\label{ns}
\bar n(s,\mathbf b, {\mathbf p}_\perp)=\frac{\sum_{n\geq 3} n\int dy I_n(s,\mathbf b, y,{\mathbf p}_\perp)}
{\sum_{n\geq 3} \int dy I_n(s,\mathbf b, y,{\mathbf p}_\perp)}
\end{equation}
and, contrary to the flows, does not affected by the unitarity correction.

   Assuming the linear dependence of the quark density
  on the distance from the cloud center and that all parts of the rotating cloud have
  the same angular velocity $\omega$, we will then obtain simple linear dependence of the mean associated
  multiplicity
  \begin{equation}\label{asm}
  \bar n (p_\perp)\simeq d(r_z)V_{r_Q}\simeq a+b p_\perp,
\end{equation}
where $d(p_\perp)$ is the density dependence on $r_z=p_\perp/\omega$, where $r_z$ is the distance
from the center of the cloud. Parameters $a$ and $b$ depend on
 the energy and impact parameter and the
parameter  $a$ should be interpreted as a quark density at the center of the constituent quark cloud.
Such linear dependence can be in fact an oversimplification, however increase of the associated multiplicity
with transverse momentum seems to be a direct consequence of the assumed constituent quark cloud rotation.
It should be noted that the $p_\perp$ and $n$ correlations
were considered as a signal for the deconfinement transition of hadronic matter long time ago by
 Van Hove\cite{vanhove}. Here we consider how the rotation effects affect such correlations.
It would be interesting to perform measurements of the associated multiplicity dependence
 on transverse momentum
and its azimuthal dependence at fixed impact parameter at RHIC and the LHC.

\section*{Discussion and conclusion}
We discussed here the nature transient state in hadronic
collisions. We believe that the same state of matter has been
revealed at RHIC in nuclei collisions.  We were concentrated on the
hadron interactions, however, we believe that the main  features  remain valid
and for nucleus interactions also,
i.e. the nature of transient state as a coherent system of
strongly interacting massive quarks is the same, its rotation
as a result of the angular momentum conservation and strong
interaction, collective effects of this rotation for the particle
production are the same too. The mechanism of  particle production
in the nuclei collisions can be different, in particular the
discussed  unitarity effects would
not play a role in the case of nuclei collisions, however the role of the
valence constituent quarks with a finite
size\footnote{We would like to speculate at this point and to mention  the possibility that the same
reason, namely the geometric size of constituent quark, can lead to the appearance of the scale
$\langle k_\perp^2\rangle \simeq 1$ $(GeV/c)^2$ in heavy quark production (cf. e.g. \cite{huang}).}\cite{morp}
 as the objects exciting the rotating
cloud of the other massive quarks seems to remain significant.
The qualitative dependence of elliptic flow for hadron collisions
is in agreement with the relevant experimental data for nuclei
collisions: increase with $p_\perp$ at small transverse momenta,
weak dependence on $p_\perp$ in the intermediate region  and
decreasing behavior with levelling off at high transverse momenta.
The new PHENIX experimental data \cite{winter} are in agreement
with this qualitative picture. It should be noted that the rotation
 effects compensate effects
of absorption and therefore the nuclear modification factor $R_{AA}$
should have a nontrivial azimuthal dependence decreasing with $\phi$.
Since the correlations are maximal
in the rotation plane, a similar dependence should be observed in the azimuthal dependence
of the two-particle correlation function.
Effect of rotation should be maximal for the peripheral collisions and therefore the
dependence on $\phi$ should be most steep at large values of impact parameter.
We would also like to stress
that linear increase with energy  of the elliptic flow in the
preasymptotic energy range  is due to increasing
density of  quarks proportional to $\sqrt{s}$ in the
transient state  which also is a reason for high parton opacity at RHIC. It
would be interesting to perform studies of transient matter at the
LHC not only in heavy ion collisions but also in $pp$--collisions
and find possible existence or absence of the rotation effects.
 Such effects should be absent if the genuine quark-gluon plasma
(gas of free quarks and gluons) would be formed at the LHC energies.

\end{document}